# GRAVITATIONAL MUSIC

**Jean-Pierre LUMINET**
*Aix-Marseille Université, CNRS, Laboratoire d'Astrophysique de Marseille (LAM) UMR 7326
& Centre de Physique Théorique de Marseille (CPT) UMR 7332
& Observatoire de Paris (LUTH) UMR 8102
France
E-mail:* jean-pierre.luminet@lam.fr

**Abstract.** Since the 1960s a growing number of composers have engaged with scientific research and have tried to incorporate their understanding of various models and theories into their musical works. Among them, Hèctor Parra (b. 1976) has been particularly impressed by the recent developments of gravitational physics and astrophysics, namely the part of astronomy which deals with gravity rather than light. Black holes, gravitational waves (first detected in September 2015), cosmology, or quantum gravity models belong to such fields of intensive research, bringing surprising new concepts such as the coarse-graining of space-time, multiverse or holographic principle. In this framework we have collaborated in the conception of a large piece for soloist ensemble, orchestra and electronic devices, which tries to transpose gravitational phenomena into a new form of contemporary music.

## 1 Science and Contemporary Music

Modern science upsets our vision of the world and influences the theory and the musical practice on an often-metaphoric mode. A recent, striking illustration was the workshop *L'influence des théories scientifiques sur le renouvellement des formes musicales dans la musique contemporaine*, held in November 2014 at the Centre de Musique Contemporaine in Paris [3]. As noticed by the organizer of the conference Mártha Grabócz, in the years 1970-1980 major advances in the field of physics, such as the catastrophe theory by R. Thom, the chaos theory by I. Prigogine, the fractal geometry by B. Mandelbrot, the big bang models of cosmology exposed in a best-seller book by S. Weinberg or the fascinating black hole studies popularized by S. Hawking, R. Penrose, J. A. Wheeler or myself, have led many composers to study scientific works and try to transpose them into musical creations [4].

As I had the privilege to give the opening lecture of the workshop [8], I began to emphasize the fact that the recent deep-sky observations by large ground-based and space telescopes, as well as by neutrino detectors, gravitational interferometers, particle accelerators, and the astrophysical modeling aimed to interpret such experimental data, revealed that space was no more the harmonious, peaceful and unchanging place dreamt by generations of scholars from Pythagoras to Kepler and Newton, but the arena of discordant, arrhythmic and utterly violent phenomena. Taking this for granted, a number of creations in contemporary music have tried to reflect this new, disturbing picture of cosmic frenzy. I gave several examples of contemporary works that used mathematical and/or astrophysical structures to create new forms of musical language. I had myself the chance to have specific collaborations with leading composers, such as Gérard Grisey for *Le Noir de l'Étoile* (1990), written for six percussionists, magnetic tape and on-site broadcast of astronomical signals emitted by pulsars [6]. Pulsars are rapidly rotating neutron stars, i.e. compact remnants of supernovae explosions that emit electromagnetic radiation in the form of periodic radio pulses. It is technically possible to monitor the pulses and figure out a way to translate them live, into sound, as the genesis of a piece of music. This was done in a striking way in Grisey's masterpiece.

The next speaker at the CDMC workshop was Hèctor Parra. The first time I had met him was in 2011, when he visited me in my office of Paris-Meudon Observatory, and I was delighted to hear that the reading of my book *Le destin de l'univers, trous noirs et énergie sombre* [7] had played a role in the composition of his work of 2011 *Caressant l'horizon* [1]. Before this, Hèctor had composed *Hypermusic Prologue* (2009), a chamber opera written in collaboration with the Harvard physicist Lisa Randall, a specialist of the so-called string theory [11]. Indeed these two pieces emerged from the keen interest of Hèctor (whose father was an engineer, as he told me later) for theoretical physics, and more particularly on the new vision of the world offered by Einstein's theory of general relativity, with such spectacular consequences as space-time curvature, time elasticity, gravity waves, black holes, wormholes or cosmological big bang models. As Hèctor explained [10], in *Caressant l'horizon* he tried to transpose the physical feeling one could have by being crossed by strong gravity waves generated by the coalescence of a binary black hole.

## 2  From black holes to quantum gravity

Science often creates its own language full of technical words, but also uses terms that belong to the layman vocabulary, although with a very different meaning that may be confusing for the non-informed reader. For instance, the nice and poetic title of Hèctor's piece *Caressant l'horizon* refers to usual terms such as "caresser" (to caress) and "horizon". However, in the vocabulary of black hole physics, "horizon" refers to the surface of a black hole (more precisely called the *event horizon*), viewed as a purely geometric border between the external universe and the inner zone of no return. And the seemingly attractive word "caress" refers to the fact that any object doomed to graze the event horizon of a black hole will suffer a violent and destructive gravitational pull, not at all a pleasant loving touch! Twenty-five years ago I myself studied how a star grazing the event horizon of a giant black hole undergoes huge tidal forces and is transformed into a stellar pancake before being fully destroyed in a titanic explosion [2]. Our telescopes have since captured such scenes.

Nevertheless, beyond such possible confusions of terms, it is important to feed the numerous gateways between science and art, even when we deal with technical concepts hardly comprehensible to the layman. Hèctor is fascinated by the relationship between the notions of a strongly distorted, curved or "atomized" space-time in the neighborhood of black hole and big bang singularities, and the nature of life itself, taking place in the apparently peaceful oasis of our planet Earth.

Einstein's theory of general relativity provides a consistent picture of the propagation of gravitation. One century ago, he discovered special solutions of the gravitational field equations representing waves of space-time curvature, travelling at the velocity of light. The major difficulty for detecting gravitational waves is their extreme weakness. The best astrophysical generators of gravitational waves are closely coupled pairs of black holes, that radiate enough gravitational energy for its effects to be directly detectable, at least in the last phase of coalescence of the binary system and the merger into a single black hole. The gravitational luminosity from the collision of two stellar mass black holes is expected to be more than the electromagnetic luminosity of the full universe filled with its hundreds of billion galaxies! However, the amplitude of the signal received on Earth is so weak that it is only in September 2015 that two detectors in the United States, achieved the required sensitivity to catch several signals of black hole coalescences, whose waveforms strikingly corresponded to the theoretical calculations previously performed by supercomputers [5]. Thus gravitational astronomy was born, opening a new window on the universe and one of the most promising developments of astrophysics for the decades to come.

Another spectacular prediction of general relativity (although not yet checked by astronomical observations) is the possible existence of "wormholes", long a staple of science fiction, as they arise as solutions to Einstein's equations describing the inner structure of rotating black holes. Such "bridges" or "throats" could join two different regions of space-time, that in some cases one is at liberty to interpret also as "parallel universes". In the upper universe, the throat has the appearance of a black hole consuming matter, light and energy; below, it appears as a white fountain expelling matter, light and energy. Thus a wormhole can be thought of as a tunnel with two ends at separate points in space-time, and the tunnel would function as a shortcut offering fantastic possibilities for interstellar travel.

Now, when the curvature radius of space-time reaches huge values, as it is the case near the center of a black hole or very close to the big bang, then matter, energy, and the gravitational field must all be quantized, and this requires a comprehensive treatment of new theories called « quantum gravity ».

Quantum mechanics governs the evolution of the microscopic world. It describes local interactions between elementary particles, which are considered to be point-like, within the geometry of a fixed background, namely, the flat space-time of Newtonian mechanics or of special relativity. Throughout the twentieth century, developments in quantum mechanics provided a unified description of electromagnetic and short-range nuclear interactions, leading to the powerful "Standard Model" of particle physics. Its technological successes are innumerable, including lasers, transistors, supercomputers, etc. At the astronomical scale, however, quantum effects play no role at all, and physics is reduced to its classical version. Gravity reigns supreme, and this is described by general relativity, a geometric theory of space-time that models gravity, not as a force acting in a space with a fixed background, but instead as the curvature of space-time, whose value at each point changes when matter is displaced. It too has shown remarkable power in describing astrophysical and cosmological phenomena.

Faced with these two domains of quite different scales and theoretical frameworks—which, through an abuse of the language, are commonly called the "infinitely small" and the "infinitely large"—the physicists may apply either quantum theory or relativity depending on whether they are dealing with small or large-scale phenomena. If this strategy is perfectly appropriate for a practical approach, it is not at all satisfactory for a theoretical one. One wants physics as a whole to be coherent. From this perspective, there must be a theory of a higher order of which quantum mechanics and general relativity are but approximations, valid only in their respective domains. This theory is called quantum gravity.

True comprehension of the Big Bang and of black holes precisely requires a theory of quantum gravity. Given the technical difficulty of the problem, it was necessary to wait until the beginning of the 1990s for physicists to learn how to transform classical theories into quantum theories. For this they needed to develop complex mathematical tools, such as knot theory, gauge group theory, topology, and so on. For a half-century, two main routes to quantifying gravity have been followed.

The first privileges the geometric approach of general relativity, and seeks thus to quantify space and time themselves. The fundamental idea is that there exist "atoms of space", namely entities at a level in which quantum effects radically change the texture of space-time itself, making it granular. The second route privileges the approach of quantum field theory, and thus considers gravitation as a force field between the objects, which it seeks to quantify in terms of an exchange of particles within a passive, exterior space-time. The fundamental ingredient is that the particles are extended objects, not points, with the radical consequence that space has more than three dimensions. These two methods have followed twisted and difficult paths. The formalism of loops for the first approach, and that of strings for the second represent their current state.

In these two cases as well, quantum gravity theories — which are still under construction — put in question the entire foundation of our understanding of the universe. That is to say, they both imply, in very different ways, the fascinating concept of a "multiverse". This term refers to a hypothetical group containing a very large, perhaps infinite, number of "possible" universes, of which ours would just be one individual example. The multiverse—the collection of all the universes in existence—would represent all that exists: space in the fullest sense of the term, time, matter, energy, and all the possible physical laws, together with the fundamental constants of nature which determine them.

Eventually, in some approaches (e.g. string theory), it is hypothesized that all the information about the matter and energy within the three-dimensional volume of a black hole is encoded as a hologram upon its two-dimensional surface, the event horizon. Such an idea, initially applied to black holes only in the hope to solve the so-called information problem, has been studied by hundreds of physicists, with hopes that it could be generalized for *any* physical system occupying a finite volume of space-time. According to this so-called "holographic principle", all the physics within a N-dimensional volume, including gravitational phenomena, could be entirely described by a simpler quantum theory without gravitation that operates only on the (N-1)-dimensional border of the volume [9].

## 3 The Inscape Project

It is fascinating to see how the most difficult and speculative concepts of modern physics, such as quantum gravity, holography or multiverse models briefly mentioned above, can be transposed into artworks. Of course it is not about strict equivalences, it has more to do with metaphors and analogies.

The work of the artist often translates the fragility of human existence, its extreme limitation in time. When we begin thinking of our existence in a cosmic perspective, we end inevitably in a kind of derisory status of human condition, so limited in space and time. Nevertheless, it is in this so short lifetime that our brain seems to be able to decipher the hidden laws of our immense universe, which normally should overtake us completely. Thus our "cosmic feeling", if it can generate some anxiety or more simply humility, may also provide some pride coming precisely from the fact that, in spite of our short-lived and infinitely modest dimension, we may succeed in achieving some big things.

Physics is precisely such a big thing, but for me another very big thing is the musical composition, that I place over any other kind of artistic creation, because it is the one which seems to me the most elaborate, the most emotional also although the most abstract, and it is probably my training in mathematics and theoretical physics that feeds my fascination for the musical language, that I had the opportunity to acquire very early, in my childhood.

Anyway, the various feelings generated by our own condition of limited human beings extend far beyond our practice as a musician or a physicist. Could "being human" just mean to be moved by the always-deeper knowledge of the surrounding nature, what we are, where we are, what we see? It seems to me that great music can give and amplify a sensation of emotional, physical and intellectual self-fulfillment, at the same time as it can give us the sensation to be overtaken. It is partly on this psychological basis that with Hèctor Parra we undertook to build together a narrative structure as a basis for an orchestral composition entitled *Inscape,* planned for 2018.

We had a first working session in January 2016, at my home place in Marseille. The musicologist José Luis Besada, a post-doc researcher at IRCAM whose project was to study *in vivo* the genesis of the work, emphasizing on the cognitive sharing and transfers between the composer and his scientific collaborators, accompanied us.

Hèctor had the idea to transpose into sounds a realistic journey through the mysteries of space-time curved by gravity, taking account of the various related physical phenomena. Another key idea was to create an acoustic illusion in which the audience would have the feeling to be in a hologram. Indeed, using the sophisticated tools of electroacoustic music, we wanted to look for a sound equivalent of a hologram, which is of course primarily visual. As holography assumes that the information contained in a volume can be entirely encoded on its border, we began to think about various ways to create acoustic holograms. This could be achieved for instance by a specific arrangement of the orchestra and space position of the instruments. We could also have musical sequences, which resume a former sequence already played by the orchestra in the full volume of the concert hall, but replayed only on the edges although displaying the same quantity of information.

How to realize that? Hèctor proposed that the acoustic superstructure played on the edge could echo the breath of the public. All the sound information generated by the spectators, their breaths, whispers, even coughing, would be encoded in some way on the edge of the concert hall - Universe.

I suggested that it could be also interesting to transpose the idea that the universe can be perceived at three very different scales, from the very small (micro scale) to the very large (macro scale), passing by the intermediate human scale (meso scale), and to make listen how we can acoustically travel through these scales.

After a night of reflection, the next morning Hèctor and I exchanged our ideas, and we noticed that we had, independently one of the other, ended in a common structure that followed a sequence of physical events associated to a gravitational journey through a giant black hole. Here are some elements of the (very tentative) scenario:

*Start from the meso scale by hearing some biological activity (human noise, whispers, etc.) with sounds confined within a small volume. Then the music grows and amplifies at the macro-scale to simulate the genesis of a massive star. At this moment an instability occurs and gravitational collapse begins. Acceleration, transposition of relativistic effects such as time dilation and space contraction, emission of gravitational waves and formation of the event horizon of a black hole. Once the black hole formed, astrophysical processes take place around: formation of a turbulent gaseous accretion disk, orthogonal jets of gases at relativistic speeds, slow growth of the black hole by accretion of matter and energy. It becomes supermassive. Destruction of full stars by tidal effects and star-star collisions: flambéed stellar pancakes, accidental supernovae. Approach and collision of two galaxies, each harboring a supermassive black hole. Ballets of their central black holes. Loss of progressive energy by emission of gravitational waves. Acceleration, orbital spiraling. Huge burst of gravitational energy when the two black holes merge into a single one. Plunging into the giant black hole. All the information of the matter and energy, which formed the black hole, is deposited on the event horizon, in the form of bits. Entering the wormhole: passage through an annular singularity, taking out of the wormhole by a white fountain. Birth of a "baby-universe" which a dodecahedral topology of space. Expansion of the new universe, at first slowed down, then accelerated by dark energy. Decreasing of dark energy, general collapse. Atomization of the space, inversion of gravitational collapse and Big Bounce.*

## 4  Work in Progress

A second working session took place in Paris one year later, in January 2017, at IRCAM, with Thomas Goepfer as the computer music director of the project. We worked in a room equipped with a system of spatialization of sound able to create the illusion of an acoustic hologram. With Thomas and Hèctor, we looked for new electronic treatments in order to transpose acoustically gravity waves and the idea of holography. Eventually Hèctor asked me to draft a text to be recorded in various ways, speeds and tones that would be distorted by electronic treatment. I provided (in French) the following text (in a very approximate English translation).

*"The star that was light became dark, silent, fathomless. Black hole, funnel of cold hells. Once crossed its horizon, an endless fall towards a bottomless center. Mixed, inverted, time and space tighten and collide. The first state of the world vaporizes into interlaced elementary grains. What become matter, energy, waves that fall? Are there a bottom, a final stop of fall, and a crushing singularity? Yet an absolute end cannot exist. An inexhaustible blooming is the only possible fruit. At the bottom of the black hole a tunnel now opens, a shortcut leading*

*to somewhere else in the universe, or even in other universes. When any border vanishes in all directions, there is no more remedy to dizziness. New universes mature, as full and delicious as newborn fruits. Our Big Bang is just the moment when occurred a bounce. From then on, the metamorphosis of the many-worlds is faster than we can think. When sprouting out of their quantum matrix, baby-universes are endowed with inconceivable shapes. The ignorant considers them flat, Nature thickens their curvatures. And more and more widely space extends, expanding, always expanding, still and forever beyond. The sky, brimming with dark energy, becomes of a frightening transparency."*

From now on, Hèctor Parra is working on the musical composition and I am eager to be in the concert hall for the world creation of *Inscape* in Barcelona on 19 May 2018, followed in June by performances in Lille and Paris.